\documentclass[12pt,thmsa]{article}

\def \eq {\begin{equation}}
\def \fim-eq {\end{equation}}
\begin{document}

\author{C. R. Carvalho$^1$ \footnote{email:crenato@if.ufrj.br}~, E. S. Guerra$^2$, Ginette Jalbert$^1$ and J. C. Garreau$^3$ \\
$^1$ Instituto de F\'{\i}sica, Universidade Federal do Rio de Janeiro \\
Cx.Postal 68528, 21941-972 Rio de Janeiro, RJ, Brazil 
\\
$^2$ Departamento de F\'{\i}sica, Universidade Federal Rural do Rio de Janeiro \\
Cx. Postal 23851, 23890-000 Serop\'edica, RJ, Brazil
\\
$^3$ Laboratoire de Physique des Lasers, Atomes et Mol\'ecules\\
Universit\'e des Sciences et Technologies de Lille\\
Bat. P5, F-59650 Villeneuve d'Ascq Cedex, France
}
\title{ Nonprobabilistic teleportation of field state via cavity QED}
\maketitle

\begin{abstract}
\noindent In this article we discuss a teleportation scheme of coherent states of cavity field. The experimental realization proposed makes use of cavity quatum electrodynamics involving the interaction of Rydberg atoms with micromaser and Ramsey cavities.
In our scheme the Ramsey cavities and the atoms play the role of auxiliary systems used to teleport the state from a micromaser cavity to another. We show that, even if the correct atomic detection fails in the first trials, one can succeed in teleportating the cavity field state if the proper measurement occurs in a later atom.

\ \newline

\end{abstract}

\section{Introduction}

Quantum teleportation was first proposed by Bennett {\it et al} \cite{Bennett} and it is a consequence of entanglement and nonlocality in quantum mechanics. These features were first noticed by Einstein, Podolsky and Rosen \cite{epr} who have originally proposed the EPR experiment in order to show that quantum mechanics were not a complete theory to describe reality. At the same time Schr\"odinger has done a formal discussion about the description and the measurements performed on two system which have interacted and that are far apart from each other \cite{schrodinger}. These strange features of quantum mechanics are consequence of the superposition principle, which leads to a quantum system to exist in a linear superposition of different eigenstates of an observable. They were the cause of an intense and long debate \cite{bell} whose result was in quantum mechanics' favour \cite{aspect}. 

Since the proposal of Bennett and co-workers for teleportation \cite{Bennett}, it has been carried out using pairs of entangled photons \cite{exptel-1}, NMR \cite{exptel-2}, as well as trapped ions \cite{exptel-3}. Besides, several schemes have been suggested to implement the teleportation in cavity QED \cite{david,Zheng,Guerra01,Guerra02}, but these schemes are probabilistic in the sense that one depends on a successful and specific sequence of measurements which one can not {\it a priori} guarantee the out coming result. In this article we propose a scheme in which flows of Rydberg atoms, properly prepared, interact with a setup, involving three superconducting cavities ($C_1$, $C_2$ and $C_3$), to teleport the field state of a cavity to another. Hence, we are concerned with the state of the system formed from the three cavities ($C_1$, $C_2$ and $C_3$). In the beginning we suppose the system in a pure state. After interacting with the first atom of the atomic beam, the global state is still a pure state. In order to teleport the field state from a cavity to another, specific measurements have to be done, i.e., one has to measure atoms in one of its possible states. The central point is: what does it happen if any detection fails? The three cavities' state becomes a statistical mixture. We show that this statistical mixture evolves in a such manner which permits one to obtain the teleportation as long as one performs the correct measurements, even with failures during the process. The only restriction concerns to the field's coherence lifetime $\tau_{coeh}$ inside the cavities: all the correct measurements have to be done in a time less than $\tau_{coeh}$.

This paper is organized as follow. In section 2 we show the process of preparing the Bell states for a system formed from two cavities, which is a necessary condition to the teleportation. We follow the basic ideas present in Ref.  \cite{crc}, which discuss schemes to entangle atoms, to achieve this goal. In section 3 we show that, through appropriate measurements, we can perform the teleportation and, finally, in section 4 we discuss the feasibility of our proposal considering the experimental limitations.

\section{Realization of Bell states}

Consider a three-level cascade atom \ $A_k$ with $\mid h_k\rangle ,\mid
e_k\rangle $ and $\mid g_k\rangle $ being the upper, intermediate and
lower atomic state respectively (see Fig. 1). We assume that the transition $
\mid h_k\rangle \rightleftharpoons \mid e_k\rangle $ is far from
resonance with the cavity central frequency such that only virtual
transitions occur between these states (only these states interact with the
cavity field). In addition we assume that the transition $\mid h_k\rangle
\rightleftharpoons \mid g_k\rangle $ is highly detuned from the cavity
frequency so that there will be no coupling with the cavity field. However, from now on we shall
consider only the levels $\mid e_k\rangle $ and $ \mid g_k\rangle$. We will not consider the level $\mid h_k\rangle$ anymore, since it will not play any role in our scheme except to introduce the phase factor in the time evolution operator (see Eq.(\ref{op-evol})) due to the 
dispersive interaction, whereas the states  $\mid e_k\rangle $ and $ \mid g_k\rangle$ are coupled in the Ramsey cavities, which we use to perform transformation in the atomic state \cite{crc}. Therefore, concerning the whole system, we have effectively a two-level system involving states $\mid e_k\rangle $ and $|g_k\rangle $.
Considering these levels, when the atoms cross the cavities $C_i$ ($i=1,2,3$), we can write an effective time evolution operator \\
\begin{equation}
U_{A_kC_i}=e^{i\varphi_1 a^{\dagger }a}\mid e_k\rangle \langle e_k\mid
+ e^{i\varphi_2 a^{\dagger }a} |g_{k}\rangle \langle g_{k}\mid ,  \label{op-evol}
\end{equation}
according to the well known interaction of a three-level atom with a single mode of the electromagnetic field (see Appendix of Ref. \cite{Guerra02}, and also Refs. \cite{eberly} and \cite{haroche}) . 

In (\ref{op-evol}) $a$ $ (a^{\dagger })$ is the annihilation (creation) operator for the field in the cavity. $\varphi_1 =\kappa_{eh}^{2}\tau /\Delta_{eh} $, \ $\kappa_{eh}$ is the coupling constant between the states $\mid h\rangle ,\mid
e\rangle $, $\Delta_{eh} =\omega _h-\omega _e-\omega $ is the detuning, where \ $\omega_h$ and $\omega _e$ \ are the frequencies of the upper and intermediate levels respectively.  $\omega $ is the cavity field frequency and $\tau $ is the atom-field interaction time. $\varphi_2 =\kappa_{eg}^{2}\tau /\Delta_{eg} $, \ $\kappa_{eg}$ is the coupling constant between the states $\mid e\rangle ,\mid g\rangle $, $\Delta_{eg} =\omega _e-\omega _g-\omega $ is the detuning \ where \ $\omega_e$ and $\omega _g$ \ are the frequencies of the intermediate and lower levels respectively. Since $\kappa_{eg}$ is smaller than $\kappa_{eh}$ and $\Delta_{eg} \ll \Delta_{eh}$, one can consider $\varphi_2 \approx 0$.  In what follows we consider $\varphi_1 =\pi$.

A sketch of the Bell states' preparation is displayed in Fig.2. Let us assume that we have a source $S_A$ which can prepare the atoms of an atomic beam in one of the states
\begin{equation}
\mid A_k^{\pm} \rangle =\frac{1}{\sqrt{2}}(\mid e_k \rangle \pm \mid g_k\rangle ).  
\label{Ak}
\end{equation}
Evidently the source $S_A$ consist of a system involving an oven, from which the atoms emerge, a velocity selection device, an excitation zone --- where the atoms are prepared in the state $|g\rangle$ --- and a classical microwave cavity, which transforms the state $|g\rangle$ in one of the states $\mid A_k^{\pm} \rangle$ \cite{brune}. Suppose that $A_1$ is the first atom coming from $S_A$ and interacting with the cavity $C_1$ prepared in a coherent state $|-\alpha \rangle _1$. Then, taking into account the evolution operator $U_{A_1C_1}$ (see Eq.(\ref{op-evol})), the system $A_1C_1$ evolves to
\begin{equation}
\mid \psi \rangle _{A_1C_1}=\frac{1}{\sqrt{2}} \Bigl( \mid e_1\rangle |\alpha \rangle _1
+ \mid g_1\rangle |-\alpha \rangle _1 \Bigr).
\end{equation}
If we define the even and odd coherent states 
\begin{eqnarray}
|{\cal{E}}_i \rangle  &=&\frac{1}{\sqrt{N_{i}^{+}}} \Bigl( |\alpha \rangle _i + |-\alpha
\rangle _i \Bigr),  \nonumber \\
|{\cal{O}}_i\rangle  &=&\frac{1}{\sqrt{N_{i}^{-}}} \Bigl( |\alpha \rangle _i - |-\alpha
\rangle _i \Bigr),  \label{eocs}
\end{eqnarray}
with $N_{i}^{\pm }=2\left( 1\pm e^{-2\mid \alpha \mid ^{2}}\right) \approx 2$
and $\langle {\cal{O}}_i\mid {\cal{E}}_i\rangle \approx 0$ \cite{alpha}, we have
\begin{equation}
\mid \psi \rangle _{A_1C_1}=\frac{1}{2} \Bigl[ |{\cal{E}}_1\rangle \Bigl( \mid e_1\rangle
+\mid g_1\rangle \Bigr)+ |{\cal{O}}_1\rangle \Bigl( \mid e_1 \rangle - \mid g_1\rangle \Bigr) \Bigr].
\end{equation}
Making use of (\ref{Ak}) we can rewrite the above expression as 
\begin{equation}
\mid \psi \rangle _{A_1C_1}=\frac{1}{\sqrt{2}}\left( |{\cal{E}}_1\rangle \mid A_1^{+} \rangle
+ |{\cal{O}}_1\rangle |A_1^{-} \rangle \right).
\end{equation}
Now we let atom $A_1$ fly through another cavity $C_2$ prepared in the
coherent state $|-\alpha \rangle _2$ and, again taking into account $U_{A_1C_2}$ (\ref{op-evol}), we have
\begin{equation}
\mid \psi \rangle _{A_1C_1C_2}=\frac{1}{2}\Bigl[ |{\cal{E}}_1\rangle \Bigl( |{\cal{E}}_2 \rangle
|A_1^{+} \rangle + |{\cal{O}}_2\rangle |A_1^{-} \rangle \Bigr) + 
|{\cal{O}}_1\rangle \Bigl( |{\cal{E}}_2\rangle |A_1^{-} \rangle + |{\cal{O}}_2\rangle
|A_1^{+} \rangle \Bigr) \Bigr].
\end{equation}
Then $A_1$ enters a Ramsey cavity $R_1$ where the atomic states are rotated according to
\begin{equation}
|A_1^{+} \rangle \longrightarrow \mid e_1\rangle ~~\mbox{  and  }~~
|A_1^{-} \rangle \longrightarrow \mid g_{1}\rangle ,
\label{Ramsey}
\end{equation}
leading the system to the state
\begin{equation}
\mid \psi \rangle _{A_1C_1C_2}=\frac{1}{2}\Bigl[ |{\cal{E}}_1\rangle \Bigl( |{\cal{E}}_2 \rangle
\mid e_1\rangle + |{\cal{O}}_2\rangle \mid g_{1}\rangle \Bigr) + 
|{\cal{O}}_1\rangle \Bigl( |{\cal{E}}_2\rangle \mid g_{1}\rangle + |{\cal{O}}_2\rangle
\mid e_1\rangle \Bigr) \Bigr]. \label{psia1c1c2}
\end{equation}
Then, after passing through the cavity $R_1$, if the detector $D_A$ measures atom $A_1$ in the level $\mid e_1\rangle $, we get for the system consisted of the two cavities $C_1$ and $C_2$ the state
\begin{equation}
|\Phi ^{+}\rangle _{C_1C_2}=\frac{1}{\sqrt{2}} \Bigl( |{\cal{E}}_1\rangle |{\cal{E}}_2\rangle
+ |{\cal{O}}_1\rangle |{\cal{O}}_2\rangle \Bigr).  \label{PHIC1C2CS+}
\end{equation}
In the last passage, if instead measuring $A_1$ in $\mid e_1\rangle $ we had measured it in the level $\mid g_1\rangle $, the state of $C_1$ and $C_2$ would have collapsed into the state 
\begin{equation}
\mid \Psi ^{+}\rangle _{C_1C_2}=\frac{1}{\sqrt{2}}\Bigl( |{\cal{E}}_1\rangle |{\cal{O}}_2\rangle + |{\cal{O}}_1 |{\cal{E}}_2\rangle \Bigr).  \label{PSIC1C2CS+}
\end{equation}

Following a similar procedure, in which the only difference is $C_2$ prepared initially in the state $|\alpha \rangle _2$, we can also prepare the states 
\begin{equation}
\mid \Phi ^{-}\rangle _{C_1C_2}=\frac{1}{\sqrt{2}} \Bigl( |{\cal{E}}_1\rangle |{\cal{E}}_2\rangle 
- |{\cal{O}}_1\rangle |{\cal{O}}_2\rangle \Bigr),  \label{PHIC1C2CS-}
\end{equation}
and
\begin{equation}
\mid \Psi ^{-}\rangle _{C_1C_2}=\frac{1}{\sqrt{2}} \Bigl( |{\cal{E}}_1\rangle |{\cal{O}}_2\rangle
 - |{\cal{O}}_1\rangle |{\cal{E}}_2\rangle \Bigr).  \label{PSIC1C2CS-}
\end{equation}
The states (\ref{PHIC1C2CS+}), (\ref{PHIC1C2CS-}), (\ref{PSIC1C2CS+}) and (
\ref{PSIC1C2CS-}) are Bell states and form a Bell basis \cite{Nielsen}.

But it is important to notice that, in order to obtain any of the states (\ref{PHIC1C2CS+}), (\ref{PHIC1C2CS-}), (\ref{PSIC1C2CS+}) and (\ref{PSIC1C2CS-}), it is necessary to measure the atom $A_1$ in a specific state, $\mid e_1\rangle$ or $\mid g_1\rangle$. Suppose we want to construct the state (\ref{PHIC1C2CS+}), i.e.,
$$
|\Phi ^{+}\rangle _{C_1C_2}=\frac{1}{\sqrt{2}} \Bigl( |{\cal{E}}_1\rangle |{\cal{E}}_2\rangle
+ |{\cal{O}}_1\rangle |{\cal{O}}_2\rangle \Bigr). 
$$
After passing through $R_1$, concerning the detector $D_A$, the atom $A_1$ has its state given by the reduced density operator
\begin{equation}
\rho_{A_1}= Tr_{C_1C_2}\Bigl[ |\psi \rangle _{A_1C_1C_2} \langle \psi |_{A_1C_1C_2} \Bigr]
= \frac{1}{2} \Bigl( |e_1\rangle \langle e_1| + |g_1\rangle \langle g_1| \Bigr), \label{rhoa1}
\end{equation}
where $|\psi \rangle _{A_1C_1C_2}$ is given by (\ref{psia1c1c2}). Therefore, considering an ideal detector adjusted for measuring the atom $A_1$ only in the state $|e_1\rangle$, we have a $50\%$ probability of failure in obtaining the above Bell state for the system $C_1C_2$. If it were an ideal detector, the failure in measuring the state $|e_1\rangle$, would correspond to measure it in the state $|g_1\rangle$. Consequently the system $C_1C_2$ would collapses into the state (\ref{PSIC1C2CS+}). In fact, we do not deal with ideal detectors and for this reason, if it does not measure the atom in the state $|e_1\rangle$, we can not say anything else than Eq.(\ref{rhoa1}). Besides, if no atom has even been registered, how can we be sure that an atom had passed through the cavity? This can be solved with a ``detector'' which is composed of two devices: a detector adjusted for measuring the atom $A_1$ only in the state $|e_1\rangle$ and an ionising chamber 
which detects the passage of the atom whatever its state.

Thus we have to take it into account when we think a proposal to construct an entangled state, such as the above Bell state (\ref{PHIC1C2CS+}), in order to perform teleportation. Our teleportation scheme, shown in section 3, was planned to work with the state (\ref{PHIC1C2CS+}), hence our first task is to guarantee we can obtain it without any chance of failure or, at least, with a good chance of success.

Then, suppose we fail in measuring $A_1$ in the state $|e_1\rangle$. The system $C_1C_2$ will be described by the reduced density operator 
\begin{eqnarray}
\rho_{C_1C_2}&=& Tr_{A_1}\Bigl[ |\psi \rangle _{A_1C_1C_2} \langle \psi |_{A_1C_1C_2} \Bigr], \nonumber \\ 
&=& \frac{1}{4} \Bigl[~ |{\cal{E}}_1{\cal{E}}_2\rangle \langle {\cal{E}}_1{\cal{E}}_2 | +
|{\cal{E}}_1{\cal{E}}_2\rangle \langle {\cal{O}}_1{\cal{O}}_2 | \nonumber \\
& & ~~~~|{\cal{E}}_1{\cal{O}}_2\rangle \langle {\cal{E}}_1{\cal{O}}_2 | +
|{\cal{E}}_1{\cal{O}}_2\rangle \langle {\cal{O}}_1{\cal{E}}_2 | \nonumber \\
& & ~~~~|{\cal{O}}_1{\cal{E}}_2\rangle \langle {\cal{O}}_1{\cal{E}}_2 | +
|{\cal{O}}_1{\cal{E}}_2\rangle \langle {\cal{E}}_1{\cal{O}}_2 | \nonumber \\
& & ~~~~ |{\cal{O}}_1{\cal{O}}_2\rangle \langle {\cal{O}}_1{\cal{O}}_2 | +
|{\cal{O}}_1{\cal{O}}_2\rangle \langle {\cal{E}}_1{\cal{E}}_2 | ~ \Bigr] ,
\label{rhoc1c2}
\end{eqnarray}
where $|\psi \rangle _{A_1C_1C_2}$ is given by (\ref{psia1c1c2}). As it can be easily  verified, $\rho_{C_1C_2}$ correspond to a statistical mixture.

The next step is to analyse the effect caused by the passage of the following atom of the atomic beam through $C_1$ and $C_2$. Call this atom $A_2$. Before entering the cavity $C_1$, the system $A_2C_1C_2$ is described by the density operator
\begin{equation}
\rho_{A_2C_1C_2}(0)=\rho_{C_1C_2}\otimes\rho_{A_2},
\label{rhoc1c2a2-0}
\end{equation}
where $\rho_{C_1C_2}$ is given by Eq.(\ref{rhoc1c2}) and 
$\rho_{A_2}=|A_2^+\rangle\langle A_2^+|$ is the corresponding density operator of the atomic pure state $|A_2^+\rangle$ (see Eq.(\ref{Ak})).

The passage of one atom $A_k$ through $C_i$ ($i=1$ or 2) is given by the operator $U_{A_kC_i}$, Eq.(\ref{op-evol}), and its result is easy to be obtained, when the system is in a pure state, as we have done above. Since the system is now in a statistical mixture described by (\ref{rhoc1c2a2-0}), the passage of one atom through one of the cavities can be described by $U_{A_kC_i}$ according to
\begin{equation}
\rho_{A_2C_1C_2}(\mbox{after})=U_{A_2C_i} \Bigl[\rho_{A_2C_1C_2}(\mbox{before}) \Bigr] U_{A_2C_i}^{\dagger} ~~\mbox{with $i$=1 or 2},
\label{evol-rho}
\end{equation}
where $\rho_{A_2C_1C_2}({\rm before})$ and $\rho_{A_2C_1C_2}({\rm after})$ are the density operator before and after the passage of atom $A_k$ through one of the cavities, respectively.
Then, after $A_2$ passes through the firts cavity $C_1$, it yields
\begin{eqnarray}
\rho_{A_2C_1C_2}(1)&=&U_{A_2C_1} \Bigl[\rho_{A_2C_1C_2}(0) \Bigr] U_{A_2C_1}^{\dagger} \nonumber \\
&=& \frac{1}{4} \Bigl[~~ |A_2^+ {\cal E}_1 {\cal E}_2\rangle\langle A_2^+ {\cal E}_1 {\cal E}_2|
- |A_2^+ {\cal E}_1 {\cal E}_2\rangle\langle A_2^- {\cal O}_1 {\cal O}_2| \nonumber \\
& & ~~~ + |A_2^+ {\cal E}_1 {\cal O}_2\rangle\langle A_2^+ {\cal E}_1 {\cal O}_2| 
- |A_2^+ {\cal E}_1 {\cal O}_2\rangle\langle A_2^- {\cal O}_1 {\cal E}_2| \nonumber \\
& & ~~~ + |A_2^- {\cal O}_1 {\cal E}_2\rangle\langle A_2^- {\cal O}_1 {\cal E}_2|
- |A_2^- {\cal O}_1 {\cal E}_2\rangle\langle A_2^+ {\cal E}_1 {\cal O}_2| \nonumber \\
& & ~~~ + |A_2^- {\cal O}_1 {\cal O}_2\rangle\langle A_2^- {\cal O}_1 {\cal O}_2| 
- |A_2^- {\cal O}_1 {\cal O}_2\rangle\langle A_2^+ {\cal E}_1 {\cal E}_2| ~~\Bigr],
\label{rhoc1c2a2-1}
\end{eqnarray}
where $\rho_{A_2C_1C_2}(0)$ is given by Eq.(\ref{rhoc1c2a2-0}). Now, after passing through $C_2$, the atom $A_2$ leaves the system described by
\begin{eqnarray}
\rho_{A_2C_1C_2}(2)&=&U_{A_2C_2} \Bigl[\rho_{A_2C_1C_2}(1) \Bigr] U_{A_2C_2}^{\dagger} \nonumber \\
&=& \frac{1}{4} \Bigl[~~ |A_2^+ {\cal E}_1 {\cal E}_2\rangle\langle A_2^+ {\cal E}_1 {\cal E}_2|
+ |A_2^+ {\cal E}_1 {\cal E}_2\rangle\langle A_2^+ {\cal O}_1 {\cal O}_2| \nonumber \\
& & ~~~ + |A_2^- {\cal E}_1 {\cal O}_2\rangle\langle A_2^- {\cal E}_1 {\cal O}_2| 
+ |A_2^- {\cal E}_1 {\cal O}_2\rangle\langle A_2^- {\cal O}_1 {\cal E}_2| \nonumber \\
& & ~~~ + |A_2^- {\cal O}_1 {\cal E}_2\rangle\langle A_2^- {\cal O}_1 {\cal E}_2|
+ |A_2^- {\cal O}_1 {\cal E}_2\rangle\langle A_2^- {\cal E}_1 {\cal O}_2| \nonumber \\
& & ~~~ + |A_2^+ {\cal O}_1 {\cal O}_2\rangle\langle A_2^+ {\cal O}_1 {\cal O}_2| 
+ |A_2^+ {\cal O}_1 {\cal O}_2\rangle\langle A_2^+ {\cal E}_1 {\cal E}_2| ~~\Bigr].
\label{rhoc1c2a2-2}
\end{eqnarray}
Then, following its trajectory, the atom $A_2$ passes through $R_1$, which perform the transformation (\ref{Ramsey}), and the system becomes
\begin{eqnarray}
\rho_{A_2C_1C_2}&=& \frac{1}{4} \Bigl[~ |e_2 {\cal E}_1 {\cal E}_2\rangle\langle e_2 {\cal E}_1 {\cal E}_2|
+ |e_2 {\cal E}_1 {\cal E}_2\rangle\langle e_2 {\cal O}_1 {\cal O}_2| \nonumber \\
& & ~~ + |g_2 {\cal E}_1 {\cal O}_2\rangle\langle g_2 {\cal E}_1 {\cal O}_2| 
+ |g_2 {\cal E}_1 {\cal O}_2\rangle\langle g_2 {\cal O}_1 {\cal E}_2| \nonumber \\
& & ~~ + |g_2 {\cal O}_1 {\cal E}_2\rangle\langle g_2 {\cal O}_1 {\cal E}_2|
+ |g_2 {\cal O}_1 {\cal E}_2\rangle\langle g_2 {\cal E}_1 {\cal O}_2| \nonumber \\
& & ~~ + |e_2 {\cal O}_1 {\cal O}_2\rangle\langle e_2 {\cal O}_1 {\cal O}_2| 
+ |e_2 {\cal O}_1 {\cal O}_2\rangle\langle e_2 {\cal E}_1 {\cal E}_2| ~\Bigr].
\label{rhoc1c2a2-3}
\end{eqnarray}
Now, if we succeed in measuring the atomic state $|e_2\rangle$, it yields
\begin{eqnarray}
\rho_{C_1C_2}&=& \frac{1}{2} \Bigl[~~ |{\cal E}_1 {\cal E}_2\rangle\langle {\cal E}_1 {\cal E}_2|
+ |{\cal E}_1 {\cal E}_2\rangle\langle {\cal O}_1 {\cal O}_2| \nonumber \\
& & ~~~ + |{\cal O}_1 {\cal O}_2\rangle\langle {\cal O}_1 {\cal O}_2| 
+ |{\cal O}_1 {\cal O}_2\rangle\langle {\cal E}_1 {\cal E}_2| ~~\Bigr],
\label{rhoc1c2-4}
\end{eqnarray}
which is the density operator corresponding to the pure state (\ref{PHIC1C2CS+})
$$ |\Phi ^{+}\rangle _{C_1C_2}=\frac{1}{\sqrt{2}} \Bigl( |{\cal{E}}_1\rangle |{\cal{E}}_2\rangle
+ |{\cal{O}}_1\rangle |{\cal{O}}_2\rangle \Bigr). $$
Otherwise, if we fail again in measuring the atomic state, the system $C_1C_2$ will be described by the partial trace of Eq.(\ref{rhoc1c2a2-3}) with respect to the atomic states, i.e,
\begin{equation}
\rho_{C_1C_2} = Tr_{A_2}\Bigl[ \rho_{A_2C_1C_2} \Bigr],
\end{equation}
whose result is
\begin{eqnarray}
\rho_{C_1C_2} &=& \frac{1}{4} \Bigl[~ |{\cal{E}}_1{\cal{E}}_2\rangle \langle {\cal{E}}_1{\cal{E}}_2 | + |{\cal{E}}_1{\cal{E}}_2\rangle \langle {\cal{O}}_1{\cal{O}}_2 | \nonumber \\
& & ~~~~|{\cal{E}}_1{\cal{O}}_2\rangle \langle {\cal{E}}_1{\cal{O}}_2 | +
|{\cal{E}}_1{\cal{O}}_2\rangle \langle {\cal{O}}_1{\cal{E}}_2 | \nonumber \\
& & ~~~~|{\cal{O}}_1{\cal{E}}_2\rangle \langle {\cal{O}}_1{\cal{E}}_2 | +
|{\cal{O}}_1{\cal{E}}_2\rangle \langle {\cal{E}}_1{\cal{O}}_2 | \nonumber \\
& & ~~~~ |{\cal{O}}_1{\cal{O}}_2\rangle \langle {\cal{O}}_1{\cal{O}}_2 | +
|{\cal{O}}_1{\cal{O}}_2\rangle \langle {\cal{E}}_1{\cal{E}}_2 | ~ \Bigr] , \nonumber
\end{eqnarray}
which is exactly Eq.(\ref{rhoc1c2})!

Therefore, in order to obtain the Bell state (\ref{PHIC1C2CS+}) for $C_1$ and $C_2$, we have only to wait the passage of the atoms until one of them is measured in the appropriate state $|e\rangle$: while the detector does not measure any atom in the state $|e\rangle$, the system $C_1C_2$ returns to its initial state (\ref{rhoc1c2}). Hence, our problem concerns the relation between the field's coherence lifetime $\tau_{coeh}$ \cite{coeh-t} in $C_1$ and $C_2$, and the time necessary to pass a large number of atoms so that the probability of measuring one of them in state $|e\rangle$ can be supposed approximately equal to one. For a cavity damping time $\tau_{cav}$ equal to $10^{-1}$ s and an average number of photons in the cavities $|\alpha|^2$ equals to $9$, the decoherence time $\tau_{coeh}=\tau_{cav}/2|\alpha|^2$ is $5 \times 10^{-3}$ s. Even with a no high detection efficiency, if we have an atomic flux of $2500$ atoms per second, it is sufficient to detect one in every ten atoms in order to guarantee the preparation of the desired Bell state. 

\section{Teleportation of a field state}

We imagine that two parties, Alice and Bob, far apart from each other, have the cavities $C_2$ and $C_1$, respectively, prepared according to last section in the state (\ref{PHIC1C2CS+}). Suppose that Alice wants to teleport the state 
\begin{equation}
|\psi\rangle_{C_3} = {\cal Y}_1 |{\cal E}_3\rangle + {\cal Y}_2 |{\cal O}_3\rangle,
\label{psi3-ini}
\end{equation}
of a third cavity $C_3$ to Bob. Thus the state of the cavity $C_1$ is to be changed to the above one(\ref{psi3-ini}). The whole experimental arrangement is sketched in Fig.3. It is consisted of five arms. The first one correspond to the setup to prepare the Bell state involving the cavities $C_1$ and $C_2$. The second one corresponds to the preparation of the state (\ref{psi3-ini}), which should be performed at the same time of the preparation of the Bell state. We let to appendix  a brief discussion about it. The third one is composed of an atomic source $S_A$, the cavities $C_2$ and $C_3$, a Ramsey cavity $R_2$ and a detector $D_A$. From $S_A$ come the atoms of type $A$ prepared in the state $|A^+\rangle$ (\ref{Ak}), which are the same used in the last section. In the fourth and fifth arms one finds the sources $S_B$, which we supposed initially turned off and from which come the atoms of type $B$. Contrary to atoms $A$, the atoms $B$ are two-level atoms resonant with the cavities $C_2$ and $C_3$ and their appropriate detection unravel $C_1$ from $C_2C_3$, leading $C_1$ to the state (\ref{psi3-ini}), as we wish. The detectors $D_A$ and $D_B$ detect the atoms $A$ and $B$, respectively.

If we let an atom $A_1$ crosses through $C_2$ and $C_3$, the whole system evolves from the state
\begin{equation}
|\psi\rangle_{C_1C_2C_3A_1}=|\Phi^+\rangle_{C_1C_2}|\psi\rangle_{C_3} |A_1^+\rangle =
\frac{1}{\sqrt{2}} \Bigl( |{\cal{E}}_1\rangle |{\cal{E}}_2\rangle + |{\cal{O}}_1\rangle |{\cal{O}}_2\rangle \Bigr) \Bigl({\cal Y}_1 |{\cal E}_3\rangle + {\cal Y}_2 |{\cal O}_3\rangle \Bigr) 
|A_1^+\rangle
\end{equation}
to the state
\begin{equation}
|\psi\rangle_{C_1C_2C_3A_1}=\frac{1}{\sqrt{2}} \Bigl[ \Bigl( {\cal Y}_1 |{\cal E}_1 {\cal E}_2 {\cal E}_3\rangle + {\cal Y}_2 |{\cal O}_1 {\cal O}_2 {\cal O}_3 \rangle \Bigr) |A_1^+\rangle - 
\Big( {\cal Y}_1 |{\cal O}_1 {\cal O}_2 {\cal E}_3\rangle + {\cal Y}_2 |{\cal E}_1 {\cal E}_2 {\cal O}_3 \rangle \Bigr) |A_1^-\rangle \Bigr]
\end{equation}
according to the evolution operator (\ref{op-evol}) applied to each cavity, $C_2$ and $C_3$, when atom $A_1$ passes through them. After the atom $A_1$ crosses the Ramsey cavity $R_2$, which performs the transformation (\ref{Ramsey})
$$
|A_1^{+} \rangle \longrightarrow \mid e_1\rangle ~~\mbox{  and  }~~
|A_1^{-} \rangle \longrightarrow \mid g_{1}\rangle ,
$$
we obtain
\begin{equation}
|\psi\rangle_{C_1C_2C_3A_1}=\frac{1}{\sqrt{2}} \Bigl[ \Bigl( {\cal Y}_1 |{\cal E}_1 {\cal E}_2 {\cal E}_3\rangle + {\cal Y}_2 |{\cal O}_1 {\cal O}_2 {\cal O}_3 \rangle \Bigr) |e_1\rangle - 
\Big( {\cal Y}_1 |{\cal O}_1 {\cal O}_2 {\cal E}_3\rangle + {\cal Y}_2 |{\cal E}_1 {\cal E}_2 {\cal O}_3 \rangle \Bigr) |g_1\rangle \Bigr].
\label{28.2}
\end{equation}
At this moment we would like to detect the atomic state $|e_1\rangle$, in order to obtain the system $C_1C_2C_3$ in the state
\begin{equation}
|\psi\rangle_{C_1C_2C_3}=\frac{1}{\sqrt{2}} \Bigl( {\cal Y}_1 |{\cal E}_1 {\cal E}_2 {\cal E}_3\rangle + {\cal Y}_2 |{\cal O}_1 {\cal O}_2 {\cal O}_3 \rangle \Bigr),
\label{32.2}
\end{equation}
but suppose we fail. Then the system $C_1C_2C_3$ is described by
\begin{eqnarray}
\rho_{C_1C_2C_3} &=& Tr_{A_1} \Bigl[ |\psi\rangle_{C_1C_2C_3A_1} \langle \psi|_{C_1C_2C_3A_1} \Bigr]
\nonumber \\
&=& \frac{1}{2} \Bigl[ ~\Bigl( {\cal Y}_1 |{\cal E}_1 {\cal E}_2 {\cal E}_3\rangle\ + {\cal Y}_2 |{\cal O}_1 {\cal O}_2 {\cal O}_3 \rangle \Bigr)\Bigl( {\cal Y}_1^* \langle{\cal E}_1 {\cal E}_2 {\cal E}_3| + {\cal Y}_2^* \langle{\cal O}_1 {\cal O}_2 {\cal O}_3 | \Bigr) 
\nonumber \\
& & + \Bigl( {\cal Y}_1 |{\cal O}_1 {\cal O}_2 {\cal E}_3\rangle + {\cal Y}_2 |{\cal E}_1 {\cal E}_2 {\cal O}_3 \rangle \Bigr)\Bigl( {\cal Y}_1^* \langle{\cal O}_1 {\cal O}_2 {\cal E}_3| + {\cal Y}_2^* \langle{\cal E}_1 {\cal E}_2 {\cal O}_3 | \Bigr) \Bigr] .
\label{29.1}
\end{eqnarray}

The effect of the next atom crossing $C_2$, $C_3$ is given by
\begin{equation}
\rho_{C_1C_2C_3A_2}=U_{A_2C_3}U_{A_2C_2}\Bigl[\rho_{C_1C_2C_3}\otimes |A_2^+\rangle\langle A_2^+| \Bigr]U_{A_2C_2}^{\dagger}U_{A_2C_3}^{\dagger},
\end{equation}
in accordance with Eq.(\ref{evol-rho}), and after passing the Ramsey cavity, it yields
\begin{eqnarray}
\rho_{C_1C_2C_3A_2} &=& \frac{1}{2} \Bigl[ ~\Bigl( {\cal Y}_1 |{\cal E}_1 {\cal E}_2 {\cal E}_3 e_2\rangle\ + {\cal Y}_2 |{\cal O}_1 {\cal O}_2 {\cal O}_3 e_2\rangle \Bigr)\Bigl( {\cal Y}_1^* \langle{\cal E}_1 {\cal E}_2 {\cal E}_3 e_2| + {\cal Y}_2^* \langle{\cal O}_1 {\cal O}_2 {\cal O}_3 e_2| \Bigr) 
\nonumber \\
& & + \Bigl( {\cal Y}_1 |{\cal O}_1 {\cal O}_2 {\cal E}_3 g_2\rangle + {\cal Y}_2 |{\cal E}_1 {\cal E}_2 {\cal O}_3 g_2\rangle \Bigr)\Bigl( {\cal Y}_1^* \langle{\cal O}_1 {\cal O}_2 {\cal E}_3 g_2| + {\cal Y}_2^* \langle{\cal E}_1 {\cal E}_2 {\cal O}_3 g_2| \Bigr) \Bigr] .
\label{31.1}
\end{eqnarray}
If now we succeed in measuring the atomic state $|e_2\rangle$, the resultant density operator of $C_1C_2C_3$ is given by 
\begin{equation}
\rho_{C_1C_2C_3}=\langle e_2|\rho_{C_1C_2C_3A_2}|e_2\rangle = \frac{1}{2} \Bigl( {\cal Y}_1 |{\cal E}_1 {\cal E}_2 {\cal E}_3 \rangle\ + {\cal Y}_2 |{\cal O}_1 {\cal O}_2 {\cal O}_3 \rangle \Bigr)\Bigl( {\cal Y}_1^* \langle{\cal E}_1 {\cal E}_2 {\cal E}_3 | + {\cal Y}_2^* \langle{\cal O}_1 {\cal O}_2 {\cal O}_3 | \Bigr) ,
\end{equation}
which correspond to the pure state (\ref{32.2}). However, if we fail again, the resultant density operator is 
\begin{equation}
\rho_{C_1C_2C_3} = Tr_{A_2} \Bigl[ \rho_{C_1C_2C_3A_2} \Bigr],
\end{equation}
where $\rho_{C_1C_2C_3A}$ is given by (\ref{31.1}), and it yields Eq. (\ref{29.1}).

Therefore, while the cavities sustain their coherence, it does not matter if we fail in measuring the atomic state $|e\rangle$ in the first, second,...trial, because the system returns always to the same configuration (\ref{29.1}) and we can try to detect until we have success.

According to the definition of the even and odd coherent state (\ref{eocs}), the state (\ref{32.2}) can be written as
\begin{eqnarray}
|\psi\rangle_{C_1C_2C_3}&=&\frac{1}{\sqrt{2}} \Bigl( {\cal Y}_1 |{\cal E}_1 {\cal E}_2 {\cal E}_3\rangle + {\cal Y}_2 |{\cal O}_1 {\cal O}_2 {\cal O}_3 \rangle \Bigr) \nonumber \\
&=& \Bigl( {\cal Y}_1 |{\cal E}_1\rangle + {\cal Y}_2 |{\cal O}_1\rangle \Bigr)\Bigl( \frac{|\alpha_2\rangle |\alpha_3\rangle + |-\alpha_2\rangle |-\alpha_3\rangle}{2}\Bigr) \nonumber \\
& & + \Bigl( {\cal Y}_1 |{\cal E}_1\rangle - {\cal Y}_2 |{\cal O}_1\rangle \Bigr)\Bigl( \frac{|\alpha_2\rangle |-\alpha_3\rangle + |-\alpha_2\rangle |\alpha_3\rangle}{2}\Bigr).
\label{32.3}
\end{eqnarray}
Then, we suppose that the detection of $|e\rangle$ turns off the atomic beam $A$ and turns on the two atomic beams $B$. This can be done with a fast eletronics, in which it is used the Stark effect to tune properly the atoms within the cavities. The atoms $B$ are two-level atoms resonant with the cavities $C_2$ and $C_3$, with $|b\rangle $ and $|a\rangle $ being their lower and upper levels, respectively. Besides, with the detection of $|e\rangle$, we also suppose that it is injected the states $|\alpha_2\rangle$ and $|\alpha_3\rangle$ in the cavities $C_2$ and $C_3$, respectively, through a classical current oscillating in antennas coupled to them. Thus, before the first two atoms, $B_2$ and $B_3$, arrive in the cavities, $C_2$ and $C_3$, respectively, the state $|\psi\rangle_{C_1C_2C_3}$ becomes
\begin{eqnarray}
|\psi\rangle_{C_1C_2C_3} &=& \Bigl( {\cal Y}_1 |{\cal E}_1\rangle + {\cal Y}_2 |{\cal O}_1\rangle \Bigr)\Bigl( \frac{|2\alpha_2\rangle |2\alpha_3\rangle + |\emptyset_2\rangle |\emptyset_3\rangle}{2}\Bigr) \nonumber \\
& & + \Bigl( {\cal Y}_1 |{\cal E}_1\rangle - {\cal Y}_2 |{\cal O}_1\rangle \Bigr)\Bigl( \frac{|2\alpha_2\rangle |\emptyset_3\rangle + |\emptyset_2\rangle |2\alpha_3\rangle}{2}\Bigr),
\label{33.1}
\end{eqnarray}
where $|\emptyset_i\rangle$ denotes the vacuum in the cavity $C_i$ ($i=2,3$). Hence, at this moment, the state of the whole system is given by $|\psi\rangle_{C_1C_2C_3B_2B_3} = |\psi\rangle_{C_1C_2C_3} \otimes |b\rangle_2 |b\rangle_3$, whose evolution, due to the atoms' passage through the cavities, is governed by the well known interaction of a two-level atom resonant with a single mode of the eletromagnetic field \cite{scully}: if atom $B_j$, initially in the lower state $|b\rangle_j$, encounters the cavity $C_j$ in the vacuum ($j=2,3$), their state does not change, it remains $|b_{j}\rangle |\emptyset_j\rangle$; however, if $B_j$ encounters the cavity in the state $|2\alpha_j \rangle$, their state evolves to 
\begin{equation}
|b_j\rangle|2\alpha_j \rangle \longrightarrow |a_{j}\rangle |\chi ^a_j\rangle + |b_{j}\rangle |\chi^b_j\rangle = |a_{j}\chi ^a_j\rangle + |b_{j}\chi^b_j\rangle,
\end{equation}
where $|\chi^a_j\rangle$ and $|\chi^b_j\rangle $ represent two different states of the field. In fact, in this case their expression are well known --
$$
|\chi^a_j\rangle =-i\sum\limits_{n}C_{n+1}\sin (gt\sqrt{n+1})|n_j\rangle \mbox{, }
|\chi^b_j\rangle=\sum\limits_{n}C_{n}\cos (gt\sqrt{n})|n_j\rangle \mbox{ and } C_{n}=e^{-\frac{1}{2}|2\alpha _j|^{2}}(2\alpha _j)^{n}/\sqrt{n!}
$$
-- but for the rest of the work $|\chi\rangle$ shall represent any possible state of the field inside one of the cavities.

Therefore, after $B_2$ and $B_3$ pass the cavities, the system state reads
\begin{eqnarray}
|\psi\rangle_{C_1C_2C_3B_2B_3}=\Bigl( {\cal Y}_1|{\cal E}_1\rangle + {\cal Y}_2|{\cal O}_1\rangle \Bigr)\Bigl[ \Bigl( |a_2\chi^a_2\rangle+|b_2\chi^b_2\rangle \Bigr)
\Bigl( |a_3\chi^a_3\rangle+|b_3\chi^b_3\rangle \Bigr) +
|b_2\emptyset_2\rangle|b_3\emptyset_3\rangle \Bigr] \nonumber \\
+ \Bigl( {\cal Y}_1|{\cal E}_1\rangle - {\cal Y}_2|{\cal O}_1\rangle \Bigr)\Bigl[ \Bigl( |a_2\chi^a_2\rangle+|b_2\chi^b_2\rangle \Bigr) |b_3\emptyset_3\rangle
+ |b_2\emptyset_2\rangle \Bigl( |a_3\chi^a_3\rangle+|b_3\chi^b_3\rangle \Bigr) \Bigr]. \nonumber \\
\label{34.1}
\end{eqnarray}
If we detect the atoms in the states $|a_2\rangle$ and $|a_3\rangle$, we reach our aim: the above state yields
\begin{equation}
|\psi\rangle_{C_1C_2C_3B_2B_3}=\Bigl( {\cal Y}_1|{\cal E}_1\rangle + {\cal Y}_2|{\cal O}_1\rangle \Bigr)|\chi^a_2\rangle |\chi^a_3\rangle |a_2\rangle |a_3\rangle ,
\label{34.1-b}
\end{equation}
which corresponds to the teleportation of the state (\ref{psi3-ini}) to cavity $C_1$.
However, let us suppose that we fail this detection, in order to show that the teleportation can still be achieved. In fact, we have just to continue trying to measure the atoms in the two secundary beams $B$. Then, if we fail in the last detection, the system $C_1C_2C_3$ is described by
\begin{equation}
\rho_{C_1C_2C_3}= Tr_{B_2}\Bigl[ Tr_{B_3}\Bigl( |\psi\rangle_{C_1C_2C_3B_2B_3}\langle\psi |_{C_1C_2C_3B_2B_3} \Bigr)\Big],
\label{34.2}
\end{equation}
where $|\psi\rangle_{C_1C_2C_3B_2B_3}$ is given by (\ref{34.1}).

In order to simplify the above expression, which is rather long, let us call  
\begin{equation}
|\phi_1\rangle = {\cal Y}_1|{\cal E}_1\rangle + {\cal Y}_2|{\cal O}_1\rangle ~~~\mbox{ and }~~|\gamma_1\rangle = {\cal Y}_1|{\cal E}_1\rangle - {\cal Y}_2|{\cal O}_1\rangle.
\label{notation-2}
\end{equation}

Thus by making use of (\ref{34.1}), Eq.(\ref{34.2}) reads
\begin{eqnarray}
\rho_{C_1C_2C_3}&=& |\phi_1 \chi^a_2 \chi^a_3\rangle\langle \phi_1 \chi^a_2 \chi^a_3| + \Bigl[ |\phi_1 \chi^b_2 \chi^a_3\rangle + |\gamma_1 \emptyset_2 \chi^a_3 \rangle \Bigr]\Bigl[ \mbox{ h.c. } \Bigr] + \Bigl[ |\phi_1 \chi^a_2 \chi^b_3\rangle + |\gamma_1 \chi^a_2\emptyset_3\rangle \Bigr]\Bigl[ \mbox{ h.c. } \Bigr] \nonumber \\
&&+ \Bigl[ |\phi_1 \chi^b_2 \chi^b_3\rangle + |\gamma_1 \chi^b_2\emptyset_3\rangle + |\phi_1 \emptyset_2\emptyset_3\rangle + |\gamma_1 \emptyset_2 \chi^b_3\rangle \Bigr] \Bigl[ \mbox{h.c. } \Bigr],
\label{38.2}
\end{eqnarray}
where h.c. means Hermetian conjugate.

Now, if we consider the next coming pair of atoms, $B_2$ and $B_3$, our system is described by $\rho_{C_1C_2C_3B_2B_3}=\rho_{C_1C_2C_3} \otimes|b_2\rangle\langle b_2| \otimes  |b_3\rangle\langle b_3| $, where $\rho_{C_1C_2C_3}$ is the above expression, and every terms of $\rho_{C_1C_2C_3B_2B_3}$ will evolve, as consequence of the passage of $B_2$ through $C_2$ and $B_3$ through $C_3$, according to the evolution of their corresponding bra's and ket's:
\begin{eqnarray}
|b_j\chi_j^a\rangle  \longrightarrow |a_j\chi_j^a\prime\rangle + |b_j\chi_j^b\prime\rangle &\mbox{or}&  \langle b_j \chi_j^a| \longrightarrow \langle a_j\chi_j^a\prime| + \langle b_j\chi_j^b\prime| ,\nonumber \\
|b_j\chi_j^b\rangle  \longrightarrow |a_j\chi_j^a\prime\prime\rangle + |b_j\chi_j^b\prime\prime\rangle &\mbox{or}& \langle b_j \chi_j^b| \longrightarrow \langle a_j\chi_j^a\prime\prime| + \langle b_j\chi_j^b\prime\prime| ,\nonumber \\
|b_j\emptyset_j\rangle  \longrightarrow |b_j\emptyset_j\rangle &\mbox{or}&
\langle b_j \emptyset_j| \longrightarrow \langle b_j \emptyset_j| .
\label{39.1}
\end{eqnarray}
Hence, according to (\ref{39.1}) $\rho_{C_1C_2C_3B_2B_3}$ yields
\begin{eqnarray}
&\rho_{C_1C_2C_3B_2B_3}\!\!\!\!&=\Biggl[~|\phi_1\rangle \Bigl(|a_2\chi_2^a\prime\rangle + |b_2\chi_2^b\prime \rangle \Bigr)\Bigl(|a_3\chi_3^a\prime \rangle + |b_3\chi_2^3\prime \rangle \Bigr) \Biggr]\Biggl[ \mbox{ h.c. } \Biggr] \nonumber \\
&&+\Biggl[~ |\phi_1\rangle \Bigl( |a_2\chi_2^a\prime\prime\rangle + |b_2\chi_2^b\prime\prime \rangle \Bigr)\Bigl( |a_3\chi_3^a\prime\rangle + |b_3\chi_3^b\prime \rangle \Bigr) + |\gamma_1\rangle |b_2\emptyset_2\rangle \Bigl( |a_3\chi_3^a\prime\rangle + |b_3\chi_3^b\prime \rangle \Bigr) \Biggr] \Biggl[ \mbox{ h.c. } \Biggr] \nonumber \\
&&+\Biggl[~ |\phi_1\rangle \Bigl( |a_2\chi_2^a\prime\rangle + |b_2\chi_2^b\prime \rangle \Bigr)\Bigl( |a_3\chi_3^a\prime\prime\rangle + |b_3\chi_3^b\prime\prime \rangle \Bigr) + |\gamma_1\rangle \Bigl( |a_2\chi_2^a\prime\rangle + |b_2\chi_2^b\prime \rangle \Bigr) |b_3\emptyset_3\rangle \Biggr] \Biggl[ \mbox{ h.c. } \Biggr] \nonumber \\
&&+\Biggl\{ |\phi_1\rangle \Bigl( |a_2\chi_2^a\prime\prime\rangle + |b_2\chi_2^b\prime\prime \rangle \Bigr)\Bigl( |a_3\chi_3^a\prime\prime\rangle + |b_3\chi_3^b\prime\prime \rangle \Bigr)
+ |\gamma_1\rangle \Bigl( |a_2\chi_2^a\prime\prime\rangle + |b_2\chi_2^b\prime\prime \rangle \Bigr) |b_3\emptyset_3\rangle \nonumber \\
&&~~ + |\phi_1\rangle |b_2\emptyset_2\rangle |b_3\emptyset_3\rangle + |\gamma_1\rangle |b_2\emptyset_2\rangle \Bigl( |a_3\chi_3^a\prime\prime\rangle + |b_3\chi_3^b\prime\prime \rangle \Bigr) \Biggr\} \Biggl\{ \mbox{ h.c. } \Biggr\}.
\label{40.1}
\end{eqnarray}

Now, what we have to verify is the following: 
i) if we measure $|a_2\rangle$ and $|a_3\rangle$, do we obtain the same result that this measurement causes in (\ref{34.1-b})?; and
ii) if we fail, i.e., if we do not succeed to measure $|a_2\rangle$ and $|a_3\rangle$ simultaneously, will we obtain a reduced density operator $\rho_{C_1C_2C_3}$ similar to (\ref{38.2})? The answer to both question is yes. Let us see then the former: if we measure $|a_2\rangle$ and $|a_3\rangle$, it means that the system is projected onto the subspace associated with the projector $|a_2a_3\rangle\langle a_2a_3|$, i.e.,
\begin{equation}
|a_2a_3\rangle\langle a_2a_3|\rho_{C_1C_2C_3B_2B_3}|a_2a_3\rangle\langle a_2a_3| = \rho_{C_1C_2C_3} \otimes |a_2a_3\rangle\langle a_2a_3|,
\end{equation}
where, by inspection of (\ref{40.1}), it is easy to see that
\begin{eqnarray}
\rho_{C_1C_2C_3}=\langle a_2a_3|\rho_{C_1C_2C_3B_2B_3}|a_2a_3\rangle &\!\!\!\!=\!\!\!\!& |\phi_1\rangle\langle \phi_1| \otimes \Bigl( |\chi^a_2\prime \chi^a_3\prime \rangle\langle \chi^a_2\prime \chi^a_3\prime | + |\chi^a_2\prime\prime \chi^a_3\prime \rangle\langle \chi^a_2\prime\prime \chi^a_3\prime | \nonumber \\
&& ~~~~~~~~~~~~~~ + |\chi^a_2\prime \chi^a_3\prime\prime \rangle\langle \chi^a_2\prime \chi^a_3\prime\prime | + |\chi^a_2\prime\prime \chi^a_3\prime\prime \rangle\langle \chi^a_2\prime\prime \chi^a_3\prime \prime| \Bigr) \nonumber \\
&=\!\!\!\!& |\phi_1\rangle\langle \phi_1| \otimes \rho_{C_2C_3}.
\label{41.3}
\end{eqnarray}
$\rho_{C_2C_3}$ is no more a pure state, as it happens in (\ref{34.1-b}), but it does not matter, because we are concerned with cavity $C_1$. Indeed, according to the above equation, we have achieved our goal: we have performed the teleportation of the state (\ref{psi3-ini}) to $C_1$, which is given by $|\phi_1\rangle = {\cal Y}_1|{\cal E}_1\rangle + {\cal Y}_2|{\cal O}_1\rangle$.

Now, let us see the other question. If we fail in detecting $|a_2\rangle$ and $|a_3\rangle$ simultaneously, our system reduces to 
\begin{eqnarray}
\rho_{C_1C_2C_3}&=&Tr_{B_2}Tr_{B_3} \Bigl[ \rho_{C_1C_2C_3B_2B_3} \Bigr] = Tr_{B_2}Tr_{B_3} \Bigl[ \rho_S \Bigr]\nonumber \\
&=& \langle a_2a_3|\rho_S|a_2a_3\rangle + \langle a_2b_3|\rho_S|a_2b_3\rangle + \langle b_2a_3|\rho_S|b_2a_3\rangle + \langle b_2b_3|\rho_S|b_2b_3\rangle ,
\label{42.1}
\end{eqnarray}
where
\begin{eqnarray}
\langle a_2a_3|\rho_S|a_2a_3\rangle &=& |\phi_1 \chi^a_2\prime \chi^a_3\prime \rangle\langle \phi_1 \chi^a_2\prime \chi^a_3\prime| + |\phi_1 \chi^a_2\prime\prime \chi^a_3\prime \rangle\langle \phi_1 \chi^a_2\prime\prime \chi^a_3\prime| \nonumber \\ \nonumber \\
&&\!\!\!\!\! + |\phi_1 \chi^a_2\prime \chi^a_3\prime\prime \rangle\langle \phi_1 \chi^a_2\prime \chi^a_3\prime\prime| + |\phi_1 \chi^a_2\prime\prime \chi^a_3\prime\prime \rangle\langle \phi_1 \chi^a_2\prime\prime \chi^a_3\prime\prime| \nonumber,
\end{eqnarray}
\
\begin{eqnarray}
\langle a_2b_3|\rho_S|a_2b_3\rangle &=& |\phi_1 \chi^a_2\prime \chi^b_3\prime \rangle\langle \phi_1 \chi^a_2\prime \chi^b_3\prime| + |\phi_1 \chi^a_2\prime\prime \chi^b_3\prime \rangle\langle \phi_1 \chi^a_2\prime\prime \chi^b_3\prime| \nonumber \\ \nonumber \\
&&\!\!\!\!\! + \Bigl[~ |\phi_1 \chi^a_2\prime \chi^b_3\prime\prime \rangle + |\gamma_1 \chi^a_2\prime \emptyset_3 \rangle ~\Bigr]\Bigl[\mbox{ h.c. }\Bigr] \nonumber \\ \nonumber \\
&&\!\!\!\!\! + \Bigl[~|\phi_1 \chi^a_2\prime\prime \chi^b_3\prime\prime \rangle + |\gamma_1 \chi^a_2\prime\prime \emptyset_3 \rangle ~\Bigr]\Bigl[\mbox{ h.c. }\Bigr] \nonumber,
\end{eqnarray}
\
\begin{eqnarray}
\langle b_2a_3|\rho_S|b_2a_3\rangle &=& |\phi_1 \chi^b_2\prime \chi^a_3\prime \rangle\langle \phi_1 \chi^b_2\prime \chi^a_3\prime| + |\phi_1 \chi^b_2\prime \chi^a_3\prime\prime \rangle\langle \phi_1 \chi^b_2\prime \chi^a_3\prime\prime| \nonumber \\ \nonumber \\
&&\!\!\!\!\! + \Bigl[~ |\phi_1 \chi^b_2\prime\prime \chi^a_3\prime \rangle + |\gamma_1 \emptyset_2\chi^a_3\prime  \rangle ~\Bigr]\Bigl[\mbox{ h.c. }\Bigr] \nonumber \\ \nonumber \\
&&\!\!\!\!\! + \Bigl[~|\phi_1 \chi^b_2\prime\prime \chi^a_3\prime\prime \rangle + |\gamma_1 \emptyset_2 \chi^a_3\prime\prime \rangle ~\Bigr]\Bigl[\mbox{ h.c. }\Bigr] \nonumber,
\end{eqnarray}
\ 
\begin{eqnarray}
\langle b_2b_3|\rho_S|b_2b_3\rangle &=& |\phi_1 \chi^b_2\prime \chi^b_3\prime \rangle\langle \phi_1 \chi^b_2\prime \chi^b_3\prime| + \Bigl[~|\phi_1 \chi^b_2\prime\prime \chi^b_3\prime\prime \rangle + |\gamma_1 \chi^b_2\prime\prime \emptyset_3 \rangle + |\phi_1 \emptyset_2 \emptyset_3 \rangle + |\gamma_1 \emptyset_2\chi^b_3\prime\prime \rangle ~\Bigr]\Bigl[\mbox{ h.c. }\Bigr]\nonumber \\ \nonumber \\
&&\!\!\!\!\! + \Bigl[~ |\phi_1 \chi^b_2\prime\prime \chi^b_3\prime \rangle + |\gamma_1 \emptyset_2\chi^b_3\prime  \rangle ~\Bigr]\Bigl[\mbox{ h.c. }\Bigr]  + \Bigl[~|\phi_1 \chi^b_2\prime \chi^b_3\prime\prime \rangle + |\gamma_1 \chi^b_2\prime \emptyset_3 \rangle ~\Bigr]\Bigl[\mbox{ h.c. }\Bigr].
\label{45.2}
\end{eqnarray}
Though the density operator $\rho_{C_1C_2C_3}$ (\ref{42.1}), consisted of the above terms, is much more complex than the one given by Eq.(\ref{38.2}), it preserves the feature of that one which is important to us: it consists in every term (ket $|...\rangle$ or bra $\langle ...|$), which constitutes the new $\rho_{C_1C_2C_3}$ (\ref{42.1}), where the state $|\gamma_1 \rangle$ appears, so do the state $|\emptyset_2\rangle$ or $|\emptyset_3\rangle$. Hence, by passing a new pair of atoms in the states $|b_2\rangle$ and $|b_3\rangle$ through $C_2$ and $C_3$, respectively, and then measuring simultaneously these atoms in the states $|a_2\rangle$ and $|a_3\rangle$, it yields a density operator $\rho_{C_1C_2C_3}$ in the same form of (\ref{41.3})
$$ |\phi_1\rangle\langle \phi_1| \otimes \rho_{C_2C_3} .$$
The only difference is that $\rho_{C_2C_3}$ is more complex than before. Besides, if we fail in measuring simultaneously the states $|a_2\rangle$ and $|a_3\rangle$, we can repeat the whole process, the expression of $\rho_{C_1C_2C_3}$ becoming even more complicated, as it happens from (\ref{38.2}) to (\ref{42.1}), but still maintaining the main feature, which consist in every term where appears the state $|\gamma_1 \rangle$ it also appears the state $|\emptyset_2\rangle$ or $|\emptyset_3\rangle$.

\section{Conclusion} 

As mentioned before our concern for the feasibility of the above experimental setup consist in the time scales involved. The cavity damping time $\tau_{cav}$ (with niobium superconducting cavities at very low temperature)  is of the order of $0.1$s \cite{haroche2}. However, our restriction in the time scale is given by the decoherence time $\tau_{coeh}=\tau_{cav}/2|\alpha|^2$, whose  value is $5\times 10^{-3}$ s for an average number of photons in the cavities $|\alpha|^2$ equals to $9$. Our proposal consist of three task to be done in sequence: the first is to prepare the Bell state, involving the cavities $C_1$ and $C_2$ (arm I of Fig. 3), and the state to be teleported in $C_3$ (arm II of Fig. 3); the second is to measure properly one atom $A$ in the arm III, in order to entangle $C_1$, $C_2$ and $C_3$; and the third is to measure properly the atoms $B$ in the arms IV and V of the setup, in order to unravel $C_1$ from $C_2$ and $C_3$, and to teleport the field state. These three steps have to be done while the fields sustain their coherence in the three cavities. Considering $\tau_{coeh} = 5 \times 10^{-3}$ s, with atomic flows of $2500$ per second, one has to perform five measurements in the proper sequence in every ten atoms in order to prepare and teleport the field state. 
Certainly this is not satisfactory. However, in the last decade, one has a considerable technological improvement in the superconducting cavities: $\tau_{cav}=2\times 10^{-6}$ s in 1994 \cite{prl1}, $\tau_{cav}=220\times 10^{-6}$ s in 1996 \cite{prl2}, $\tau_{cav}=1\times 10^{-3}$ s in 2001 \cite{prl3} and $\tau_{cav}=1\times 10^{-1}$ s in 2006 \cite{haroche2}. Therefore, we believe that a cavity with a damping time of the order of $1$ s should be achieved in the next years, making our proposal reasonable, i.e., to perform five measurements in the proper sequence in every hundred atoms, instead of ten.

Concluding, we have presented a scheme of teleportation of field state. Our scheme is a deterministic one in the sense that we do not depend on a sequence of measurements, which all have to be successful at the first trial. Our proposal setup supports failures and, according to the above discussion, we believe that it has a good chance of working in the near future.

\section{Acknowledgments}
This work was partially supported by the Brazilian agencies CNPq, FUJB and
FAPERJ.

\ \newpage

\appendix

\section{Preparation of the state (\ref{psi3-ini})}
Suppose we have the cavity $C3$ prepared initially in a coherent state $|-\alpha \rangle _{3}$ and an atomic source from which come atoms prepared in the state
\begin{equation}
\mid \psi \rangle _{A}=c_{e}\mid e\rangle +c_{g}\mid g\rangle .
\end{equation}%
After the first atom flies through $C3$, taking into account the time
evolution operator (\ref{op-evol}), the state of the system $C_3A$ is given by 
\begin{equation}
|\psi \rangle _{C_3A}=c_{e}| e\rangle |\alpha \rangle_{3} + c_{g}|g\rangle |-\alpha \rangle _{3.}
\end{equation}
Then, $A$ passes through the Ramsey zone $R_3$, where the atomic states are rotated according to 
\begin{equation}
R_{3}=\frac{1}{\sqrt{2}}\left[ 
\begin{array}{cc}
1 & -ie^{i\theta } \\ 
-ie^{-i\theta } & 1%
\end{array}
\right] ,
\end{equation}
that is,
\begin{eqnarray}
|f\rangle &\rightarrow& \frac{1}{\sqrt{2}}( |e\rangle -ie^{-i\theta
}\mid g\rangle ),  \nonumber \\
|g\rangle &\rightarrow& \frac{1}{\sqrt{2}}(-ie^{i\theta }|e\rangle
+ |g\rangle ),
\end{eqnarray}
and therefore, the state of the system $C_3A$ will be given by
\begin{eqnarray}
|\psi \rangle _{C_3A} &=&\frac{1}{\sqrt{2}}\Bigl[(c_{e}-ie^{i\theta }c_{g}) |{\cal E}_3\rangle + 
(c_{e}+ie^{i\theta }c_{g}) |{\cal O}_3\rangle \Bigr]|e\rangle  \nonumber \\
&+& \frac{1}{\sqrt{2}}\Bigl[(-ie^{-i\theta }c_{e}+c_{g})| {\cal E}_3\rangle)
+ (-ie^{-i\theta }c_{e}-c_{g})| {\cal O}_3\rangle \Bigr]| g\rangle .
\end{eqnarray}
Now, in order to obtain the state $|\psi \rangle _{C3}$ in cavity $C3$, we measure the atomic state. If we detect $|e\rangle $, we have 
$${\cal Y}_1 =(c_{e}-ie^{i\theta }c_{g})/\sqrt{2} ~~\mbox{ and }~~ {\cal Y}_2 =(c_{e}+ie^{i\theta }c_{g})/\sqrt{2}.$$ Otherwise, if we detect $| g\rangle $, we have $${\cal Y}_1 =(-ie^{-i\theta }c_{e}+c_{g})/\sqrt{2} ~~\mbox{ and }~~ {\cal Y}_2 =(-ie^{-i\theta }c_{f}-c_{g})/\sqrt{2}.$$

\ \newpage

\textbf{Figure Captions} \newline

\textbf{Fig. 1} Energy states scheme of a three-level atom, where $|h\rangle$ is the upper state with atomic frequency $\omega _h$, $\ |e\rangle $ is
the intermediate state with atomic frequency $\omega _e$, $|g\rangle $ is
the lower state with atomic frequency $\omega _g$ and $\omega $ is the
cavity field frequency and $\Delta =(\omega _h-\omega _e)-\omega $ is
the detuning. The transition $\mid e\rangle \rightleftharpoons \mid h\rangle 
$ is far from resonance with the cavity central frequency such that
only virtual transitions occur between these levels (only these states
interact with field in cavity $C_i$ ($i=1,2,3$)). In addition we assume that the
transitions $|h\rangle \rightleftharpoons |g\rangle $ and $|e\rangle \rightleftharpoons |g\rangle $ are highly detuned from the cavity frequency so that there will be no coupling with the cavity field in $C_i$ ($i=1,2,3$).\newline

\textbf{Fig. 2} Sketch of the Bell states´ preparation. It involves a beam of Rydberg atoms  prepared in a source $S_A$, crossing two high-Q cavities $C_1$ and $C_2$, in which coherent states is previously injected, a low-Q cavity $R_1$, in which a classical microwave field
can be applied, and being measuring by a detector $D_A$.

\textbf{Fig. 3} Sketch of the whole experiment.


\begin{thebibliography}{99}
\bibitem{alpha} For $|\alpha|^2 = 9$, one has $e^{-2\mid \alpha \mid ^{2}} \approx 1 \times 10^{-8}$, which is in good agreement with experimental precision, and consequently $N_{i}^{\pm }=2\left( 1\pm e^{-2\mid \alpha \mid ^{2}}\right) = 2$ and $\langle {\cal{O}}_i\mid {\cal{E}}_i\rangle = 0$ are good approximations.
\end{thebibliography}
\end{document}